# Harms in Repurposing Real-World Sensory Cues for Mixed Reality: A Causal Perspective

Yujie Tao
yjtao@stanford.edu
Stanford University
Stanford, CA, USA

Sean Follmer
sfollmer@stanford.edu
Stanford University
Stanford, CA, USA

## ABSTRACT

The rise of Mixed Reality (MR) stimulates new interactive techniques that seamlessly blend the virtual and physical environments. Just as virtual content could be overlayed onto the physical world for providing adaptive user interfaces [5, 8], emergent techniques "repurpose" everyday environments and sensory cues to support the virtual content [7, 9, 13–15]. For instance, a strong wind gust in the real world, rather than being distracting to the virtual experience, can be mapped with trees swaying in MR to achieve a unifying experience [15], as shown in Figure 1. Such techniques introduce stronger immersion, but they also expose users to overlooked perceptual manipulations, where safety risks arise from misperception of real-world events. In this work, we apply a causal inference perspective to understand the harms of repurposing real-world sensory cues for MR. We argue that by viewing the MR experience as a causal inference process of interpreting cues arising from both the virtual and physical world, MR designers and researchers can gain a new lens to understand potential perceptual manipulation harms.

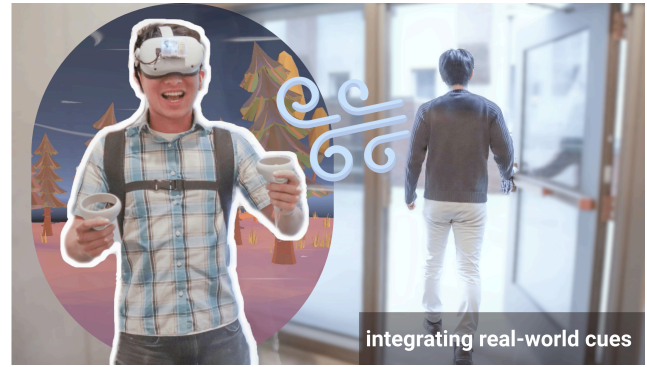

Figure 1: A demonstration of repurposing real-world sensory cues for MR. The wind gust from the real world, instead of distracting the user, is mapped with the sway of trees in MR. Figure adapted from Tao and Lopes [15].

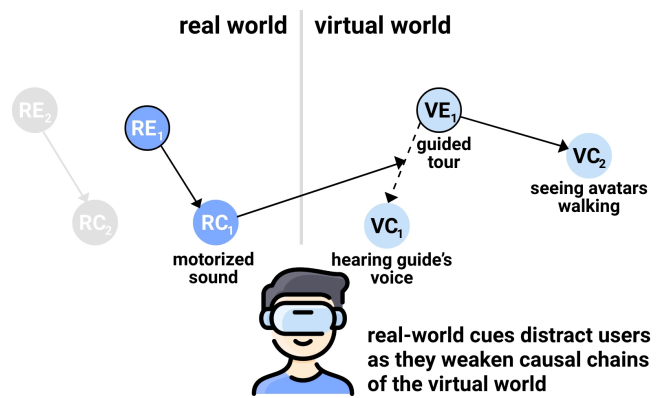

Figure 2: A conceptual causal model of perceiving multisensory cues in a MR experience. $VC$ and $VE$ denote sensory cues and events in the virtual world. $RC$ and $RE$ denotes sensory cues and events in the real world. The directional edges between nodes indicate a causal chain. The dashed lines indicate a weakened causal relationship.

## 1 A CAUSAL PERSPECTIVE

Humans understand the world by combining multisensory cues and inferring causal relationships [10]. Imagine you return home and see fragments of porcelain plates on the kitchen floor. Almost immediately, a familiar meow cuts through the silence. The synthesis of these sensory cues leads to a direct inference of the event, in which the cat's actions might have resulted in the broken plates.

Using this perspective, we can view the MR experience as the user receiving sensory cues from both the virtual and physical world, each from its own sequence of events within causal chains, as shown in Figure 2. There, the user tries to infer if their actions cause the cues they observe and remains in presence in the virtual world if the events and cues are causally congruent [3].

In the work from Tao and Lopes [15], the authors described a VR museum experience. There, the user roamed through the museum and joined a guided tour ($VE_1$), during which the user listened to the guide's narration ($VC_1$) and observed virtual avatars moving around ($VC_2$). The integration of those multisensory cues leads to immersion in the virtual museum [3].

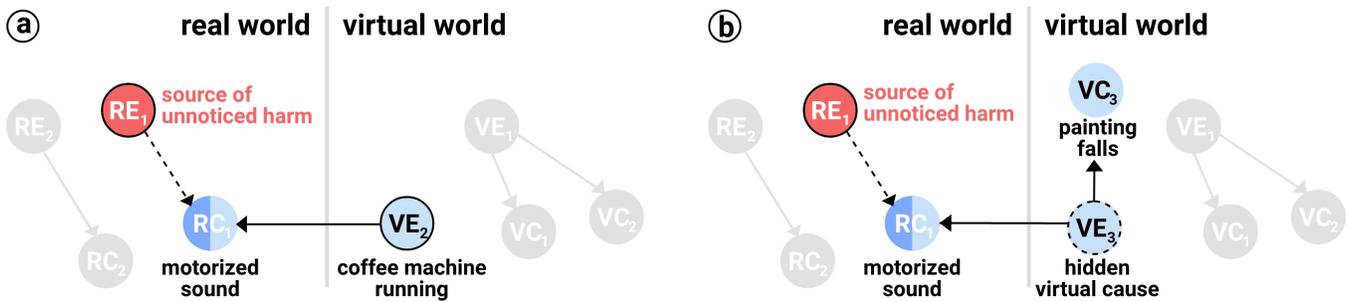

Figure 3: A conceptual causal model of repurposing real-world sensory cues for MR by (a) creating a virtual event that can be inferred as the direct cause of the real-world sensory cue; and (b) creating a virtual cue that is congruent with the real-world sensory cue. With both cues, user infers about the hidden virtual event that causes them. Safety risk emerges when user re-establishes the causal chains, as the sources of harms in the physical world now get unnoticed.

Suddenly, a long "motorized" sound emerged and broke the presence. Here, the motorized sound ($RC_1$), coming from the real world, is not congruent with any virtual cues. It weakened the causal chain of the virtual world, and the user started to infer the real-world event ($RE_1$) that caused that cue ($RC_1$).

To mitigate distractions from the real world, researchers proposed to integrate real-world sensory cues into the virtual environment by mapping them with congruent virtual content [15]. One way is to introduce a virtual event that can lead to the real-world cue. For instance, the motorized sound can be mapped with a running coffee machine in the virtual museum [15]. This technique creates a virtual event $VE_2$ (i.e., coffee machine running) that establishes a causal capture [11] with the real-world cue $RC_1$ (i.e., motorized sound), as shown in Figure 3(a).

The direct cause of an effect does not need to be observed by the user for them to make an inference of an event. Thus, another way to map the real-world sensory cue is to introduce an additional virtual cue that is congruent with the real-world sensory cue. In this example, the motorized sound can be mapped with the fall of a painting. Here, using both real-world cue ($RC_1$) and the new virtual cue ($VC_3$), the user remains immersed as they infer the hidden virtual event ($VE_3$) that causes these cues, as shown in Figure 3(b).

The potential for harm arises at this exact moment when the user re-establishes the causal chains among different sensory cues and events. Here, the motorized sound becomes an effect of both real-world and virtual-world events. The introduced causal capture weakens the evidence for the real-world event ($RE_1$), as shown in Figure 3. The motorized sound could be an effect of a wide range of events in the real world, from a coffee machine in the room to a drone flying toward the user. Depending on the nature of the real-world sources, repurposing everyday sensory cues might lead to different outcomes. When it's implemented improperly, the user misperceives real-world harm and indulges in virtual immersion at the cost of physical safety.

Prior research on causal perception can also inform researchers of new methods [1, 12] and tools [2, 6] for mitigating the harms in repurposing real-world sensory cues for MR. From a causal perspective, one way to mitigate harm is to enable the user to re-establish the causal chains of real-world events and cues when physical harms emerge. This can be instantiated by exposing the user to the sources of real-world sensory cues. For instance, by enabling notifications and/or a direct see-through of the real world, the user can re-establish the causal capture of real-world events and cues and act against the harms accordingly.

Another direction is to weaken the causal binding between the virtual events/cues and real-world cues. An instantiation of this is to adjust the virtual content so that it is no longer perceived as the cause of real-world sensory cues. For instance, by manipulating the timing, spatial location, contextual priors, and realism of the added virtual event/cue, designers can create experience where the user is less likely to perceive the real-world sensory cues to be part of the virtual environment.

## 2 FUTURE OUTLOOK

In this work, we focus on a causal inference perspective to understand the harms of repurposing real-world sensory cues for MR. Learning from causal perception in psychology, researchers can construct a causal representation of the world given the set of observations the user has. Using a causal inference framework, researchers can further computationally model the integration of multisensory cues in an MR experience using probabilistic programming [1] and hierarchical bayesian models of causal perception [12]. Such a computational model could help VR designers understand at which moment the user starts to misperceive real-world events that might expose them to higher safety risks. We believe a causal framework will inspire new mitigation techniques, optimizing for metrics that can break or enhance the causality perception when repurposing real-world sensory cues for MR. In the meantime, the probabilistic inference approach needs to be validated through human subject studies.

More broadly speaking, this work taps into understanding the risks of perceptual manipulation attacks, which is becoming an increasing concern in the use of MR. Past work on perceptual manipulation attacks identified safety risks through speculative designs [16] and interviews [4], providing qualitative understanding. Going beyond empirical approaches, we argue that constructing computational models of human sensory processes can help more comprehensively identify perceptual manipulation risks in MR.